\documentclass[english,letterpaper,aps,prb,twocolumn,amsmath,showpacs,amssymb]{revtex4}
\usepackage[T1]{fontenc}
\usepackage[latin1]{inputenc}
\usepackage{graphicx}
\usepackage{amssymb}

\makeatletter

\usepackage{epsfig}

\usepackage{dcolumn}
\usepackage{bm}
\usepackage{bbm}

\usepackage{wasysym}

\usepackage{marvosym}

\usepackage{times}
\usepackage{amsfonts}

\newlength{\textwidthm}

\usepackage{babel}
\makeatother

\begin{document}

\title{Lattice Green's function approach to the solution of the spectrum
of an array of quantum dots and its linear conductance}

\author{N.~M.~R.~Peres and T.~Stauber}

\affiliation{Centro de F\'{\i}sica and Departamento de F\'{\i}sica, Universidade
do Minho, P-4710-057, Braga, Portugal}

\author{J.~M.~B. Lopes dos Santos}

\affiliation{CFP and Departamento de F{\'{\i}}sica, Faculdade de Ci\^encias Universidade
de Porto, 4169-007 Porto, Portugal}

\date{\today}

\begin{abstract}
In this paper we derive general relations for the band-structure of
an array of quantum dots and compute its transport properties when
connected to two perfect leads. The exact lattice Green's functions
for the perfect array and with an attached adatom are derived. The
expressions for the linear conductance for the perfect array as well as
for the array with a defect are presented. The calculations are illustrated
for a dot made of three atoms. The results derived here are also the
starting point to include the effect of electron-electron and electron-phonon
interactions on the transport properties of quantum dot arrays. Different derivations of the exact lattice Green's functions are discussed. 
\end{abstract}


\maketitle

\section{Introduction}

Quantum dots, quantum wires, and molecular structures are among the
most studied low-dimensional condensed matter systems due to their
importance to nanoelectronics,\cite{dots} and recently to
biology.\cite{biology} In the field of molecular electronics,
molecules are used to control the current flow when they are assembled
in between metal contacts.  \cite{molecular} Among the several
physical properties exhibited by low-dimensional electron systems,
interesting ones can be found in the transport properties of these
systems, such as conductance quantization and conductance
oscillations,\cite{kwapinski} the latter effect depending on the
number of atoms in the wire. Some of these effects have been found in
the related field of carbon nanotubes, where experiments have shown
that the conductance of a single wall carbon nanotube is
quantized\cite{liang} and shows Fabry-Perot interference patterns, a
signature of coherent transport in the carbon wire.

The above systems all fall under the study of low-dimensional physics
and are best described starting from the tight-binding approximation.
When we address the physical properties of these low-dimensional
systems there are two different questions to address. One is concerned
with the electronic spectrum,\cite{onipko1} and the other with the
transport properties.\cite{onipko2} The first of these properties is
of importance for the optical response of the system and the second
for its use in nanoscopic devices.

The need of an efficient method of computing both the spectrum and the
conductance of these systems is therefore obvious.  Both these
problems can be traced back to the calculation of the lattice Green's
function of the system. In the case of the spectrum, the Green's
function should be computed for the isolated system, whereas when
considering the transport properties, the Green's function must be
computed taking into account the effect of the coupling between the
system and the metallic leads. The lattice Green's function method was
used to describe the appearance of surface modes (Tamm states) in a
finite one-dimensional chain and its interaction with a non-linear
impurity. \cite{molina} The generalization to a semi-infinite square
lattice was also discussed. \cite{molina2} Several different methods
of computing transport properties of one-dimensional quantum wires,
using lattice Green's functions, are available in the literature. A
tutorial overview on some of the used methods was recently written by
Ryndyk \textit{et al.}.\cite{ryndyk} Using the Keldysh method, the
coherent transport through a one-dimensional lattice was studied by
Zeng \textit{et al.}.\cite{zeng} Li \textit{et al.} studied the
transport through a quantum dot ring with four sites.\cite{li} The
inclusion of time dependent potentials on the transport properties of
one-dimensional chains was done by Arrachea. \cite{arrachea} The
inclusion of electron-electron interactions in the transport
properties of a small system was discussed by Oguri\cite{Oguri} and of
a quantum wire was done by Karrach \textit{et al.},\cite{karrach}
using a functional renormalization group method. The extension to a
quasi-one-dimensional Kagom\'e wire, where the feature of a multiband
system is present, was considered by Ishii and Nakayama.\cite{ishii}
The interesting situation where the metallic wire is connected to a
Heisenberg chain was studied by Reininghaus \textit{et
al.}. \cite{reininghaus} The above results are just a very small
subset of the existing representative literature on quantum transport,
where the concept of lattice Green's function plays a central role.

Historically, the first approach to electronic transport in a one-dimensional
finite system was done in a series of elegant papers published by
Caroli \textit{et al.}. \cite{caroli,combescot,combescot&schreder}
In these works the authors addressed the question of how defects affect
the charge transport of the quantum wire. Indeed, the question of
how localized defects change the otherwise perfect transport properties
of the system has been addressed by several authors in the framework
of tight-binding systems. \cite{Guinea87,Sautet88,Mizes91}

Guinea and Verg\'es \cite{Guinea87} used a Green's function method
to study the local density of states and the localization length of
a one-dimensional chain coupled to small pieces of a polymer. They
showed that at the band center there is a complete suppression of the
transmission coefficient due to a local antiresonance. Sautet and
Joachim \cite{Sautet88} studied the effect of a single impurity
on the transport properties of a one-dimensional chain. The impurity
was assumed to change both the on-site energy and the hopping to the
next-neighbor atoms. Mizes and Conwell \cite{Mizes91} considered
the effect of a single impurity in two coupled-chains, showing that
a change on the on-site energy has a more pronounced effect in reducing
the transmission in the one-dimensional chain than it has in this
system. Finally, Peres and Sols studied analytically the effect of
a localized defect in the transport properties of polyacene (a multi-band
system), putting in evidence a parity effect, that was also used by
Akhmerov \textit{et al.}\cite{beenakker} to formulate a theory of
the valley-valve effect in graphene nanoribbons.\cite{peres0,peres1,castro}
Also the study of vacancies in transport of quasi-two-dimensional systems
is an important area of research.\cite{wakabaiashi}
The study of a linear array of quantum dots, represented by a single
site with an $s-$orbital was carried out by Teng \textit{et al.}.\cite{teng}

The book by Ferry and Goodnick has a very good introductory section to
lattice Green's functions.\cite{ferry} Also the recursive Green's function
method is there presented. The method we present in this paper has
strong similarities to the recursive Green's function method.\cite{ferry}
The important difference to stress here is that the recursive Green's
function method is implemented as a numerical method, whereas our
approach does solve the same type of problems in an analytical
way. The link between Green's functions and transport properties of
nanostructures and mesoscopic systems is well covered in the book by
Datta.\cite{datta}

In this work we give a detailed account of a method, based on the
solution of the Dyson equation, to compute the lattice Green's function
of an array of quantum dots, where the dot is represented by an arbitrary
number of sites with arbitrary values of the site energies and of
the hopping parameters. The method is developed within the approximation
that there is only one hopping channel between the dots in the array.
This constrain is used to keep the level of the formalism at its minimum.
The formalism is easily generalized to include the effect of defects
(both on-site and adatoms defects) and to describe the transport properties
of both the clean and the perturbed system. The method exploits the
fact that the Dyson equation for the lattice Green's function can
be solved exactly for bilinear problems as long as the hoppings are
not of arbitrary long range. Also the developed formalism can be used
to describe surface states and non-linear impurities effects\cite{molina}
in a finite array of quantum dots, but we will not pursue these two
aspects in this paper. Although developed within a quasi-one-dimensional
perspective, we will show in a forthcoming publication how the method
can be generalized to two-dimensional ribbons.


\section{Determination of the lattice Green's function}

In this section we develop a general method for determining the Green's
function of an array of quantum dots. From it both the electronic
spectrum and the rule for momentum quantization are obtained. As a
warming up we first revisit the solution of the finite chain problem.

The traditional approach\cite{economou,barton,katsura} to determine
the Green's function in real space requires the previous solution
of the Schr\"odinger equation, with the corresponding determination
of its eigenvalues and eigenvectors. After this is done, the Green's
function is computed using \begin{equation}
G(\mathbf{r},\mathbf{r}',z)=\sum_{n}\frac{\psi_{n}^{\ast}(\mathbf{r})\psi_{n}(\mathbf{r}')}{z-\lambda_{n}}
\label{G}
\end{equation}
 where $\psi_{n}(\mathbf{r})$ and $\lambda_{n}$ are, respectively,
the eigenstates and eigenvalues of the eigenproblem $H\psi_{n}(\mathbf{r})=\lambda_{n}\psi_{n}(\mathbf{r})$,
where $H$ stands for the Hamiltonian of the problem and the summation
over $n$ means a discrete sum (for discrete eigenvalues),or an integration
(for continuous eigenvalues), or both, over the quantum numbers of
the problem.

The evaluation
of the sum in Eq. (\ref{G}) may be a very hard task, depending on
the mathematical complexity of both the wave functions and the eigenvalues.
Even for relatively simple cases the evaluation of the summation is
far from obvious (see appendices of Ref. \onlinecite{lindenberg}).

An alternative approach, used for lattice systems, starts from the
definition of the resolvent operator \begin{equation}
\hat{G}=\frac{1}{z-H}\,,
\label{resolv}
\end{equation}
 and computes the matrix elements of the resolvent by evaluating directly
a number of determinants associated with the matrix $(z-H)$. \cite{karadov,teng}
This method has the obvious drawback of being limited by the possibility
of computing analytically the necessary determinants. \cite{vein,dy}

We present in what follows a method that overcomes the technical difficulties
above mentioned.


\subsection{The single chain case\label{sub:The-single-chain}}

\label{schain}

In order to understand how the method works, we revisit the problem
of determining the Green's function of a finite one-dimensional chain
of atoms, with a single orbital per atom. This simple example will
help us to fix the notation and state the general arguments about
the solution of this type of problems. Let us assume that the system
has $N$ atoms, with the motion of the electrons described by the
tight-binding Hamiltonian

\begin{equation}
H=H_{0}+V\,,\label{H1D}\end{equation}
 with \begin{equation}
H_{0}=\epsilon_{0}\sum_{i=1}^{N}\vert i\rangle\langle i\vert\,,\label{H0}\end{equation}
 and \begin{equation}
V=-t\sum_{i=1}^{N-1}(\vert i\rangle\langle i+1\vert+\vert i+1\rangle\langle i\vert)\,.\label{V}\end{equation}
 Clearly, Eq. (\ref{H0}) represents the on-site energy of the electrons
in the atoms and Eq. (\ref{V}) represents the hopping of the electrons
between neighboring atoms. This may seem as an important restriction,
but it can in fact be relaxed and the approach extended to more
general hopping processes. \cite{loly} Alternatively the Green's
function of a more complex Hamiltonian may be generated using the
\textit{extension theory} for lattice Green's functions. \cite{schwalm}

Let us now introduce two different resolvent operators, the free resolvent
\begin{equation}
\hat{G}^{0}=\frac{1}{z-H_{0}}\,,\end{equation}
 and the full resolvent $\hat{G}$, given by Eq. (\ref{resolv}) with $H$
given by Eq. (\ref{H1D}). The strategy is to determine $\hat{G}$
by solving exactly Dyson's  equation,\cite{economou} considering
the hopping term $V$ as perturbation. In terms of the resolvents
and of $V$, the Dyson equation takes the form \begin{equation}
\hat{G}=\hat{G}^{0}+\hat{G}^{0}V\hat{G}\,.\label{dyson}\end{equation}
 Forming matrix elements with the basis vectors and using $\hat{\mathbf{1}}=
\sum_{i=1}^{N}\vert i\rangle\langle i\vert$,
one obtains \begin{equation}
\frac{1}{G^{0}}\langle i\vert\hat{G}\vert j\rangle=\delta_{ij}-t(\langle i-1\vert\hat{G}\vert j\rangle+\langle i+1\vert\hat{G}\vert j\rangle)\,,\label{matrixG}\end{equation}
 with $G^{0}=(z-\epsilon_{0})^{-1}$\,. Apart from the term $\delta_{i,j}$,
this is the equation for the wavefunction of a particle of coordinate
$i$ with the tight binding Hamiltonian (\ref{H1D}), and eigenvalue
$G^{0}(z)$. It is obvious that the difference of two solutions of this
equation will be a solution of the equation without the $\delta_{i,j}$
term. So $G_{i,j}\equiv\langle i\vert\hat{G}\vert j\rangle$ can be
determined by adding a general solution of the homogeneous equation
to \emph{one }particular solution of the full equation. The latter
can then be determined by the boundary conditions. 

The solutions of the homogeneous equation are superpositions of plane
waves, $\langle n\vert\hat{G}\vert m\rangle=A_{m}e^{i\theta n},$ where
$A_{m}$ is an arbitrary function of $m$, and $\theta$ is defined
by \begin{equation}
\frac{1}{G^{0}}=-2t\cos\theta\,,\label{Gteta}\end{equation}
the usual dispersion relation for a 1D tight-binding problem with
nearest neighbor hopping. To find one particular solution of the full
equation, we use the fact that it has to satisfy the homogeneous equation
for $i<j$ and $i>j$ and therefore be a linear combination of plane
waves of wavevectors $\pm\theta$---the solutions of Eq.(\ref{Gteta})---or,
equivalently, of $\sin(i\theta)$ and $\cos(i\theta)$:\begin{subequations}
\label{allequations2} \begin{eqnarray}
G_{ij}^{<}=A^{j}\cos(i\theta)+B^{j}\sin(i\theta),\textrm{for}\hspace{.5cm}i\leq j\,,\label{Gminor}\\
G_{ij}^{>}=C^{j}\cos(i\theta)+D^{j}\sin(i\theta),\textrm{for}\hspace{.5cm}i>j\,.\label{Gmajor}\end{eqnarray}
 \end{subequations}In the linear system of equations obtained by
fixing $j$ in Eq. (\ref{matrixG}) , \begin{equation}
\left[\begin{array}{cccccc}
1 & tG^{0} & 0 & \cdots & \cdots & 0\\
tG^{0} & 1 & tG^{0} & 0 & \cdots & 0\\
0 & tG^{0} & 1 & tG^{0} & \cdots & 0\\
\vdots & \ddots & \ddots & \ddots & \ddots & \vdots\\
\vdots & \ddots & \ddots & \ddots & \ddots & \vdots\\
0 & \cdots & \cdots & \cdots & tG^{0} & 1\end{array}\right]\left[\begin{array}{c}
\langle1\vert\hat{G}\vert j\rangle\\
\langle2\vert\hat{G}\vert j\rangle\\
\vdots\\
\langle j\vert\hat{G}\vert j\rangle\\
\vdots\\
\langle N\vert\hat{G}\vert j\rangle\end{array}\right]=\left[\begin{array}{c}
0\\
0\\
\vdots\\
G^{0}\\
\vdots\\
0\end{array}\right]\label{ls}\end{equation}
all but two equations are automatically satisfied by the fact that
$G_{ij}^{<}$ and $G_{ij}^{>}$ solve the homogeneous equation, leaving
only two conditions mixing $G_{ij}^{<}$ with $G_{ij}^{>}$:
\begin{subequations}
\begin{eqnarray}
\frac{1}{G^{0}}G_{jj}^{<} & = & 1-t\left(G_{j+1j}^{>}+G_{j-1j}^{<}\right)\,,\\
\frac{1}{G^{0}}G_{j+1j}^{>} & = & -t\left(G_{j+2j}^{>}+G_{jj}^{<}\right).
\end{eqnarray}
\end{subequations}
These are easily shown to be equivalent (using the fact that the Green's function is also a solution
of the homogeneous equations) to

\begin{subequations}\label{bc} \begin{eqnarray}
G_{jj}^{>} & = & G_{jj}^{<}\,,\label{eq:bc1}\\
G_{j+1j}^{>}-G_{j+1j}^{<} & = & \frac{1}{t}\label{eq:bc2}\,.
\end{eqnarray}
\end{subequations} 
The Equation (\ref{eq:bc1}) corresponds to the continuity of the Green's function,
whereas Eq. (\ref{eq:bc2}) corresponds to the discontinuity of the derivative
of the Green's function in the theory of second order differential equations.
\cite{barton}

Inserting Eqs.(\ref{allequations2}) in Eqs.(\ref{bc})
, one can obtain a rather simple solution (valid in both domains $i\leq j$
and $i>j$) as\begin{equation}
G_{ij}=\frac{1}{2t\sin\theta}\sin\left(\theta\left|i-j\right|\right)\label{eq:particular_sol}\end{equation}
The general solution is obtained by adding an arbitrary solution of
the homogeneous equation, \begin{equation}
G_{ij}=A^{j}\cos(\theta i)+B^{j}\sin(\theta i)+\frac{1}{2t\sin\theta}\sin\left(\theta\left|i-j\right|\right).\label{eq:general_sol}\end{equation}
The free coefficients, $A^{j}$ and $B^{j}$, are determined by boundary
conditions. For a finite chain we must enforce, 

\begin{subequations} \label{allequations}\begin{eqnarray}
\langle0\vert\hat{G}\vert j\rangle=\langle N+1\vert\hat{G}\vert j\rangle & = & 0\label{c1}\\
\langle i\vert\hat{G}\vert0\rangle=\langle i\vert\hat{G}\vert N+1\rangle & = & 0.\label{c2}\end{eqnarray}
 \end{subequations}It is straightforward to show that these lead
to \begin{eqnarray}
G_{ij}(z) & = & \frac{1}{2t}\frac{\cos(N\theta+\theta)}{\sin(N\theta+\theta)\sin\theta}[\cos(i\theta-j\theta)-\cos(i\theta+j\theta)]\nonumber \\
 & - & \frac{1}{2t}\left[\frac{\sin(i\theta+j\theta)}{\sin\theta}-\frac{\sin\vert i\theta-j\theta\vert}{\sin\theta}\right]\,.\label{Gsol}\end{eqnarray}
Noticing that the Chebyshev polynomials obey the finite differences equation
\begin{equation}
 f_{n-1}(x)+f_{n+1}(x)=2xf_n(x)
\end{equation}
and have the representation\cite{abramowitz}
\begin{subequations} \label{allequations3} \begin{eqnarray}
T_{n}(\cos\theta) & = & \cos(n\theta)\,,\\
U_{n}(\cos\theta) & = & \frac{\sin(n\theta+\theta)}{\sin\theta}\,,\end{eqnarray}
 \end{subequations} we see that our solution (\ref{Gsol}) is the
same obtained for a finite one-dimensional harmonic lattice, \cite{bass}
as it should be. It is worth noticing that the rule for momentum quantization
(that is $\theta$) is obtained from the poles of the Green's function
(\ref{Gsol}), leading to \begin{equation}
\sin(N+1)\theta=0\Rightarrow\theta_{\ell}=\frac{\pi{\ell}}{N+1}\,,\end{equation}
 with ${\ell}=0,1,2,\ldots,N-1$.

Rewriting $G^{0}=(z-\epsilon_{0})^{-1}$ as $G^{0}=(E-\epsilon_{0}+i0^{+})^{-1}$,
where $E$ stands for the energy of the electron, the energy dependence
of the Green's function is obtained;\cite{economou} this allows
the calculation of the local density of states from \begin{equation}
\rho_{ii}(E)=-\frac{1}{\pi}\Im G_{ii}(E+i0^{+})\,.\end{equation}
 It is clear that $\rho_{ii}(E)$ is site dependent for a finite chain.
It is an elementary calculation to find an explicit expression for
$\rho_{ii}(E)$, by writing both $\cos(2i\theta)$ and $\sin(2i\theta)$
in terms of $\sin\theta$ and $\cos\theta$ and using Eq. (\ref{Gteta}).
We conclude by stressing that the matrix elements of the resolvent
were obtained without the need of evaluating the sum over the eigenstates,
as in Eq. (\ref{G}), which constitutes the major advantages of this
method over most of the existing ones.

As a final comment in this section, we remark that the final solution
for $G_{ij}$ is quite symmetrical in both coordinates, even though
the method we followed treats the two coordinates on a rather different
footing. In Appendixes \ref{apJLS} and \ref{apTS} we give two other derivations of the
Green's function of a finite chain which treats both coordinates equally
from the beginning. 


\subsection{The general case}

\begin{figure}[ht]
 \includegraphics[width=7cm]{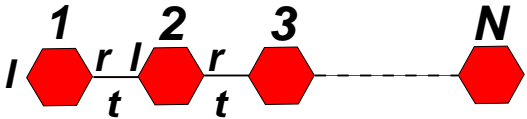} 

\caption{\label{Fig_array} An array of $N$ quantum dots coupled together
by an hopping integral $t$. The dot is represented by the hexagon.
The points $l$ and $r$ are those that are coupled between dots due
to $t$, and the choice of the letters stand for most \textit{left}
and most \textit{right} points in the dot, respectively.}

\end{figure}

We now consider the case where $N$ quantum dots, represented by hexagons
in Fig. \ref{Fig_array}, are coupled together by an hopping parameter
$t$. Although having just a single hopping channel between the dots
may seem rather restrictive, we used it nevertheless to illustrate
the method. Also, if we choose to position the dot in such a way that
is has oriented edges such that a particular site of the dot is closer
to the next dot than any other point, as in the case of Fig. \ref{Fig_array},
the used approximation is somewhat justified. The generalization of
this possibility to multi-hopping hopping processes does not changes
the general idea, but adds some complexity to the final solution.
In addition, we assume that the dot is also described by a lattice
model. Although this an apparent restriction, it can also be relaxed.
We now address the question of determining the Green's function of
the quantum dot array.

The Hamiltonian for the problem defined by Fig. \ref{Fig_array} is
written as \begin{equation}
H_{T}=H^{D}+V\,,\label{HD}\end{equation}
 were \begin{equation}
H^{D}=\sum_{i=1}^{N}H_{i}^{D}\,,\end{equation}
 with $H_{i}^{D}$ the Hamiltonian of the $i$ dot (which is not necessary
to specify at this point), and \begin{equation}
V=-t\sum_{i=1}^{N-1}(\vert i,r\rangle\langle i+1,l\vert+\vert i+1,l\rangle\langle i,r\vert)\,.\end{equation}
 As before we define $\hat{G}^{0}=(z-H^{D})^{-1}$, and note that
$\vert i,\beta\rangle$ are the basis states with $i=1,2,\ldots,N$
and $\beta$ labeling the sites in the quantum dot; we will need two
different $\beta$'s only, $\beta=r,l$.

The matrix elements of the Dyson's equation (\ref{matrixG}), given
the Hamiltonian (\ref{HD}), reads \begin{eqnarray}
\langle i,\alpha\vert\hat{G}\vert j,\beta\rangle=\delta_{i,j}G_{\alpha\beta}^{0} & - & t(G_{\alpha l}^{0}\langle i-1,r\vert\hat{G}\vert j,\beta\rangle\nonumber \\
 & + & G_{\alpha r}^{0}\langle i+1,l\vert\hat{G}\vert j,\beta\rangle)\,,\label{GHD}\end{eqnarray}
 with $G_{\alpha\beta}^{0}=\langle i,\alpha\vert\hat{G}^{0}\vert i,\beta\rangle$.
It is easy to see, by direct replacement, that the homogeneous Dyson's
equation (that is when $i\ne j$) is solved by an Ansatz of the form
$G_{\alpha\beta}^{nm}=A_{\alpha\beta}^{m}e^{in\theta}$ when $\theta$
is chosen such that 
\begin{equation}
1+2tG_{lr}^{0}\cos\theta+t^{2}(G_{lr}^{0})^{2}=t^{2}G_{ll}^{0}G_{rr}^{0}\,,\label{band}
\end{equation}
 an expression that gives the band structure once the quantized values
of $\theta$ have been determined. 
We see from Eq. (\ref{GHD}) that $\langle i,\alpha\vert\hat{G}\vert j,\beta\rangle$
is only coupled to $\langle i-1,r\vert\hat{G}\vert j,\beta\rangle$
and $\langle i+1,l\vert\hat{G}\vert j,\beta\rangle$. Therefore we
will solve the Dyson's equation (\ref{GHD}) for the particular case
of $\alpha,\beta=l,r$. To start with we make the Ansatz (a linear combination of terms of the form
$A_{\alpha\beta}^{m}e^{in\theta}$)

\begin{subequations} \begin{eqnarray}
G_{\alpha\beta}^{<,ij}=A_{\alpha\beta}^{<}\cos(i\theta)+B_{\alpha\beta}^{<}\sin(i\theta),\textrm{for}\hspace{.5cm}i<j\,,\label{GHDminor}\\
G_{\alpha\beta}^{>,ij}=A_{\alpha\beta}^{>}\cos(i\theta)+B_{\alpha\beta}^{>}\sin(i\theta),\textrm{for}\hspace{.5cm}i>j\,,\label{GHDmajor}\end{eqnarray}
 \end{subequations} where $G_{\alpha\beta}^{\lessgtr,ij}=\langle i,\alpha\vert\hat{G}\vert j,\beta\rangle$,
and the multiplicative coefficients of the trigonometric functions
depending on $j$. The finiteness of the chain is imposed by the conditions
\begin{equation}
\langle0,r\vert\hat{G}\vert j,\beta\rangle=\langle N+1,l\vert\hat{G}\vert j,\beta\rangle=0\,,\end{equation}
 and time reversal symmetry implies that \begin{equation}
\langle i,\alpha\vert\hat{G}\vert j,\beta\rangle=\langle j,\beta\vert\hat{G}\vert i,\alpha\rangle\,.\end{equation}
The solution of the non-homogeneous
Dyson's equations (that is, when $i=j$) leads to a linear system
of equations of the form \begin{equation}
M\mathbf{v}=\mathbf{b}\,,\label{Mvb}\end{equation}
 with the transpose vectors given by $\mathbf{v}^{T}=[A_{ll}^{>},A_{lr}^{>},B_{rr}^{<}]$
$\mathbf{b}^{T}=[G_{ll}^{0},G_{rr}^{0},0]$\,. The matrix $M$ is
easily constructed from the non-homogeneous Dyson's equations and
is given in the Appendix \ref{apI}. The solution of the linear system
is best accomplished by Cramer's rule, leading to \begin{subequations}
\begin{eqnarray}
A_{ll}^{>} & = & -\frac{G_{ll}^{0}}{P_{L}}\sin(N\theta+\theta)\tilde{P}_{j-1}\,,\\
A_{lr}^{>} & = & \frac{tG_{ll}^{0}G_{rr}^{0}}{P_{L}}\sin(N\theta+\theta)\sin(j\theta)\,,\\
B_{rr}^{<} & = & -\frac{G_{rr}^{0}}{P_{L}}\tilde{P}_{N-j}\,,\end{eqnarray}
 \end{subequations} with $P_{L}=tG_{lr}^{0}\sin\theta\tilde{P}_{L}$
and $\tilde{P}_{x}=tG_{lr}^{0}\sin(x\theta-\theta)+\sin(x\theta)$.

Combining the homogeneous and non-homogeneous equations we can derive the following results
\begin{subequations}
\begin{eqnarray}
G^{<,m-1m}_{rl}-G^{>,m-1m}_{rl}=t^{-1}\,,\nonumber\\
G^{>,m+1m}_{lr}-G^{<,m+1m}_{lr}=t^{-1}\,,
\label{GderivI}
\end{eqnarray}
\end{subequations}
which combined with time reversal symmetry and one of the non-homogeneous equations leads to the linear
system
\begin{equation}
 V\mathbf{u}=\mathbf{q}\,,
\label{N}
\end{equation}
with $\mathbf{u}^T=[A^>_{rl},B^>_{rl},B^<_{lr}]$, 
and $V$ and $\mathbf{q}$ given in Appendix \ref{apI}. The solution of the linear system
gives
\begin{subequations}
\begin{eqnarray}
A^>_{rl}&=& (tP_L)^{-1}\tilde P_L\tilde P_{m-1}\,,\\
B^>_{rl}&=&-(tP_L)^{-1}[\cos(L\theta)+tG^0_{lr}\cos(L\theta-\theta)]\tilde P_{m-1}\,,\\
B^<_{lr}&=&(tP_L)^{-1}[\cos\theta+tG^0_{lr}\cos(2\theta)]\tilde P_{L-m}\,.
\end{eqnarray}
\end{subequations}

As we saw before the poles of the Green's function gives the rule
of momentum quantization. In this case the poles correspond to the
zeros of $P_{L}$, which leads to an equation for the quantization
of $\theta$ that reads \begin{equation}
tG_{lr}^{0}\sin(N\theta-\theta)+\sin(N\theta)=0\,.\label{tquant}\end{equation}
 Contrary to the case of the single finite chain, Eq. (\ref{tquant})
depends on the energy, and therefore it has to be solved together
with Eq. (\ref{band}). The knowledge of $A_{ll}^{>}$, $A_{lr}^{>}$,
$A^>_{rl}$, $B^>_{rl}$, $B^<_{lr}$,
and $B_{rr}^{<}$ is all that is necessary to determine all the Green's
functions for this problem. The full form of the Green's functions
is given in Appendix \ref{apII}. 

\subsection{Explicit results for a particular example}

\label{example}

Let us now consider a specific example and work out the energy spectrum
and the momentum quantization. For the Hamiltonian of the quantum
dot we consider the case where the dot is made of three sites very
close together, with a single local orbital\cite{teng} associated
to each site. The sites are coupled together by a hopping matrix element
$t_{\triangle}$. The simpler case of representing the dot by a single
site was considered by Teng \textit{et al.}. \cite{teng} The Hamiltonian
of the dot we are considering reads \begin{eqnarray}
H_{i}^{D}= &  & -t_{\triangle}\sum_{\alpha=1}^{3}(\vert i,\alpha\rangle\langle i,\alpha+1\vert+\vert i,\alpha+1\rangle\langle i,\alpha\vert)\nonumber \\
 &  & +\epsilon_{0}\sum_{\alpha=1}^{3}\vert i,\alpha\rangle\langle i,\alpha\vert\,,\label{HDi}\end{eqnarray}
 with the boundary condition $\langle i,4\vert=\langle i,1\vert$\,.
What is now necessary is to compute the Green's function for this
system. It just happens that the Green's function for this system
is given by the Green's function of a finite chain (three sites) with
periodic boundary conditions. This can be obtained from the procedure
of Sec. \ref{schain} by replacing the boundary conditions (\ref{c1})
by \begin{equation}
\langle1\vert\hat{G}\vert j\rangle=\langle N+1\vert\hat{G}\vert j\rangle\,,\end{equation}
 which leads to \begin{eqnarray}
G_{ij}= &  & -\frac{1}{2t_{\triangle}}\frac{\sin(N\theta)}{\cos(N\theta)\sin\theta-\sin\theta}T_{i-j}(\cos\theta)\nonumber \\
 &  & +\frac{1}{2t_{\triangle}}U_{\vert i-j\vert-1}(\cos\theta)\,.\label{Gperiodic}\end{eqnarray}
 For the Hamiltonian (\ref{HDi}) we have $N=3$ and the Green's functions
computed from (\ref{Gperiodic}) are given by \begin{subequations}
\label{Gtriangle} \begin{eqnarray}
G_{rr}^{0}=G_{ll}^{0}=G_{11}^{0} & = & \frac{1}{2t_{\triangle}}\frac{1+2\lambda}{-1+\lambda+2\lambda^{2}}\,,\\
G_{lr}^{0}=G_{rl}^{0}=G_{12}^{0} & = & \frac{1}{2t_{\triangle}}\frac{1}{1-\lambda-2\lambda^{2}}\,,\end{eqnarray}
 \end{subequations} with $\lambda=(E-\epsilon_{0})/(2t_{\triangle})$.
We should note that $G_{11}^{0}$ can be written as \begin{equation}
G_{11}^{0}=\frac{1}{2t_{\triangle}}\left(\frac{2}{3}\frac{1}{\lambda-1/2}+\frac{1}{3}\frac{1}{\lambda+1}\right)\,,\end{equation}
 which means that the eigenvalue $\lambda=-1/2$ is bidegenerate,
and that is the fundamental reason why the denominator of the Green's
function is not a cubic polynomial.

\begin{figure}[ht]
 \includegraphics[clip,width=7cm]{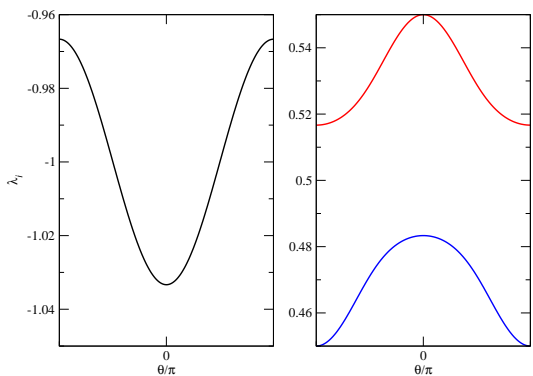} 

\caption{\label{Fig_bands} Eigenvalues $\lambda_{i}$, when $i=1$ (left)
and $i=2,3$ (right) as function of $\theta/\pi$, for large values
of $N$. We have used the ratio $t/t_{\triangle}$=0.1.}

\end{figure}

We shall now consider the physical relevant case where $t_{\triangle}\gg t$.
Within this approximation the solutions of Eq. (\ref{band}) are given
by ($\lambda\ne1$)

\begin{subequations} \label{lambda} \begin{eqnarray}
\lambda_{1} & \simeq & -1-\frac{t}{3t_{\triangle}}\cos\theta\,,\\
\lambda_{2} & \simeq & \frac{1}{2}+\frac{t\cos\theta}{6t_{\triangle}}-\frac{t}{6t_{\triangle}}\sqrt{3+\cos^{2}\theta}\,,\\
\lambda_{3} & \simeq & \frac{1}{2}+\frac{t\cos\theta}{6t_{\triangle}}+\frac{t}{6t_{\triangle}}\sqrt{3+\cos^{2}\theta}\,,\end{eqnarray}
 \end{subequations} The values of $\theta$ are obtained from the
solution of Eq. (\ref{tquant}), which requires the knowledge of $G_{12}^{0}$,
which in the approximation of Eq. (\ref{lambda}) are given by

\begin{subequations} \label{lambdaG12} \begin{eqnarray}
G_{12}^{0}(\lambda_{1}) & \simeq & -\frac{t^{-1}}{2\cos\theta}\,,\\
G_{12}^{0}(\lambda_{2}) & \simeq & \frac{t^{-1}}{-\cos\theta+\sqrt{3+\cos^{2}\theta}}\,\\
G_{12}^{0}(\lambda_{3}) & \simeq & -\frac{t^{-1}}{\cos\theta+\sqrt{3+\cos^{2}\theta}}\,\end{eqnarray}
 \end{subequations} The Eq. (\ref{tquant}) is trivially solved for
the case $\lambda=\lambda_{1}$, giving

\begin{equation}
\theta_{\ell}=\frac{\pi\ell}{N+1}\,,\end{equation}
 with $\ell=0,\pm1,\pm2,\ldots,N$. The other values of $\theta$
for $\lambda_{2}$ (top sign in Eq. (\ref{tl12})) and $\lambda_{3}$
(bottom sign in Eq. (\ref{tl12})) are obtained as solutions of \begin{equation}
\cos(N\theta)\sin\theta=\pm\sin(N\theta)\sqrt{3+\cos^{2}\theta}\,,\label{tl12}\end{equation}
 which for large $N$ reduces to \begin{equation}
\theta_{\ell}\simeq\frac{\pi\ell}{N}\,,\end{equation}
 with $\ell=0,\pm1,\pm2,\ldots,N$. This result can be appreciate
graphically, by plotting both sides of Eq. (\ref{tl12}) on the same
graph. Naturally, when $N\rightarrow\infty$, $\theta$ becomes a
continuous variable in the interval $\theta\in[-\pi,\pi[$. In Figure
\ref{Fig_bands} we plot the eigenvalues $\lambda_{i}$, given by
Eq. (\ref{lambda}), using $t/t_{\triangle}=0.1$. The three sites
composing the dot originate three energy mini-bands (since $t_{\triangle}\gg t$)
in the dot array. %

It should be now clear that this method allowed us to determine the
energy spectrum, the quantization rule for $\theta$, in the case
of a finite $N$, and the Green's functions for the array of quantum
dots, with essentially the same effort it would take us to solve the
Schr\"odinger equation for this problem.


\section{The transport properties}

We next want to work out the transport properties of the finite system
described in Sec. \ref{example}. We first derive the general results
and latter use them to study the system introduced in Sec. \ref{example}.
We will consider that our system is connected to two semi-infinite
perfect leads. The leads to which we will connect the system will
be described in similar terms to those used in the Newns model \cite{newns,davison}
and the connection between the Green's function and the transmission
across the system is described using a formalism similar to that introduced
by Mujica \textit{et al.},\cite{mujicaI,mujica,mujicabook} which,
it turns out, it is similar to the approach developed by Fisher and
Lee. \cite{fisher} The general relation between the transfer matrix
approach and the Green's function method is described in Ref. [\onlinecite{alvarez}],
in the context of continuous models.


\subsection{The general formalism}

\label{gform}

Next we present the general formalism including some particular important
results, associated with the concept of surface Green's function.
We represent the Hamiltonian of the perfect leads by one-dimensional
semi-infinite tight-binding models reading \begin{subequations} \label{Hleads}
\begin{eqnarray}
H_{L} & = & -\beta_{L}\sum_{i=-\infty}^{0}(\vert i-1,L\rangle\langle i,L\vert+\vert i,L\rangle\langle i-1,L\vert)\,,\\
H_{R} & = & -\beta_{R}\sum_{i=N+1}^{\infty}(\vert i+1,R\rangle\langle i,R\vert+\vert i,R\rangle\langle i+1,R\vert)\,,\end{eqnarray}
 \end{subequations} where $\beta_{L,R}$ are the hopping parameters
of the left ($L$) and right ($R$) leads. Although we are representing
the leads by a one-dimensional model, this is of no consequences in
the characterization of the transport properties of the dots, being
only essential that $\beta_{L,R}$ are such that metal bands in the
leads have very large band-width. Since the effective band width of
our dot structure is proportional to $t$, the only condition is that
$\beta_{L,R}\gg t$. The coupling between the leads and the dots is
made by the Hamiltonians \begin{subequations} \label{couplingleads}
\begin{eqnarray}
V_{L} & = & t_{L}(\vert0,L\rangle\langle1,l\vert+\vert1,l\rangle\langle0,L\vert)\,,\\
V_{R} & = & t_{R}(\vert N+1,R\rangle\langle N,r\vert+\vert N,r\rangle\langle N+1,R\vert)\,,\end{eqnarray}
 \end{subequations} where $t_{L,R}$ are the hopping parameters coupling
the left ($L$) and right ($R$) leads to the array of quantum dots.
We are neglecting the possibility of direct coupling between the left
and right leads, a simplification of no physical consequences, corresponding
to the fact that the array of dots has many of these. The approach
we are formulating with Eqs. (\ref{Hleads}) and (\ref{couplingleads})
differs somewhat from that of Refs. [\onlinecite{newns,mujicaI}],
but in general terms the two approaches are perfectly equivalent.

The important aspect of the tunneling approach proposed in Ref. {[}\onlinecite{mujicaI}]
is the need to compute the off-diagonal Green's function $G_{lr}^{1N}(z)$
for accessing the tunneling properties of the system, including in
the calculation the coupling to the semi-infinite leads. This can
be done in many different ways [\onlinecite{mujicaI,teng,ando91,khomyakov05,peressols,nikolic}],
leading in the end to results similar to those obtained using non-equilibrium
Green's function methods. \cite{jauho,rammer}

The full Hamiltonian of the problem is the sum of Eqs. (\ref{HD}),
(\ref{Hleads}), and (\ref{couplingleads}). Among the several ways
available to compute $G_{lr}^{1N}(z)$, one possibility is to use
again the Dyson's equation approach. This requires that we know the
exact Green's function of the leads, before the coupling to the system
is established, this is we need to compute the Green's function of
the problem defined by Eq. (\ref{Hleads}). The calculation of the
Green's function of the leads can be done using the same method we
used in Sec. \ref{schain}, but now with the boundary conditions \begin{equation}
\langle1\vert\hat{G}_{L}\vert m\rangle=\langle N\vert\hat{G}_{R}\vert m\rangle=0\,.\end{equation}
 In this case, however, it is much easier to obtain the Green's function
from the usual definition (\ref{G}). The wave function of the electrons
is given by (let us consider the left lead only) \begin{equation}
\vert\theta\rangle=\sqrt{\frac{2}{N_{L}+1}}\sum_{n=-N_{L}}^{0}\sin[(n-1)\theta]\vert n\rangle\,,\label{wf}\end{equation}
 where $N_{L}$ is a normalization \textit{length} that is taken to
infinity in the end of the calculation. Using the definition (\ref{G})
and the wave function (\ref{wf}) the matrix elements of the resolvent
is given by the integral in the complex plane (where $C$ defines
a contour over the unit circle $w=e^{i\theta}$) \begin{equation}
G_{nm,L}(z)=\frac{1}{2\pi i}\oint_{C}\frac{w^{\vert n-m\vert}-w^{\vert m+n-2\vert}}{\beta_{L}w^{2}+zw+\beta_{L}}dw\,.\label{Gsemii}\end{equation}
 The integral in (\ref{Gsemii}) can be evaluated using the same method
it has been used to solve for the Green's function of a chain with
periodic boundary conditions \cite{economou}, leading to \begin{equation}
G_{nm,L}^{ret}(E)=\frac{1}{2i\beta_{L}}(1-x^{2})^{-1/2}\left(w_{2}^{\vert n-m\vert}-w_{2}^{\vert m+n-2\vert}\right)\,,\label{GsemiiretL}\end{equation}
 with $G_{nm,L}^{ret}(E)$ standing for the retarded Green's function,
$w_{2}=-x+i\sqrt{1-x^{2}}$ and $x=E/(2\beta_{L})$, such that $\vert x\vert<1$.
The result (\ref{GsemiiretL}) is a generalization of the particular
results given in Ref. {[}\onlinecite{bishop}] {[}the same is true
for Eq. (\ref{GsemiiretR})]. Repeating the same arguments for the
right lead we obtain \begin{equation}
G_{nm,R}^{ret}(E)=\frac{1}{2i\beta_{R}}(1-x^{2})^{-1/2}\left(w_{2}^{\vert n-m\vert}-w_{2}^{\vert m+n-2N\vert}\right)\,,\label{GsemiiretR}\end{equation}
 and $x=E/(2\beta_{R})$. Central to our study are the surface Green's
functions $G_{00,L}^{ret}(E)$ and $G_{N+1N+1,R}^{ret}(E)$ which
we obtain from Eqs. (\ref{GsemiiretL}) and (\ref{GsemiiretR}).

In order to determine $G_{lr}^{1N}(z)$ we introduce $H^{0}=H_{L}+H_{T}+H_{R}$
and the full Hamiltonian $H=H^{0}+V_{L}+V_{R}$, the \textit{free}
resolvent is $\hat{G}^{0}=(z-H^{0})^{-1}$. As before, $G_{lr}^{1N}(z)$
is determined solving the Dyson's equation, which due to the short
range nature of $V_{L}$ and $V_{R}$ has an analytical solution,
reading \begin{equation}
G_{lr}^{1N}(z)=\langle1l\vert\hat{G}^{0}\vert Nr\rangle D^{-1}\,,\label{G1N}\end{equation}
 with \begin{eqnarray}
D & = & (1-t_{L}^{2}\langle1l\vert\hat{G}^{0}\vert1l\rangle G_{00,L})\nonumber \\
 & \times & (1-t_{R}^{2}\langle Nr\vert\hat{G}^{0}\vert Nr\rangle G_{N+1N+1,R})\nonumber \\
 & - & t_{L}^{2}t_{R}^{2}\langle1l\vert\hat{G}^{0}\vert Nr\rangle\langle Nr\vert\hat{G}^{0}\vert1l\rangle G_{00,L}G_{N+1N+1,R}\,.\label{D}\end{eqnarray}
 A formally equivalent result to Eq. (\ref{G1N}) was first derived
by Caroli \textit{et al.}, \cite{caroli} in the context of tunneling
across a one-dimensional wire, which was the first application of
the Keldysh\cite{rammer} formalism to tunneling problems.

Let us now describe briefly the calculation of the linear conductance,
for which Eq. (\ref{G1N}) is needed. The central quantity in this
approach is the $T-$matrix. Starting from the Dyson's equation (\ref{dyson})
and introducing an iterative solution, we arrive at an equivalent
form for $\hat{G}$, given by

\begin{equation}
\hat{G}=\hat{G}^{0}+\hat{G}^{0}\hat{T}\hat{G}^{0}\,,\end{equation}
 where $\hat{T}$ is the $T-$matrix given by \begin{equation}
\hat{T}=V+V\hat{G}^{0}\hat{T}=V+V\hat{G}V\,,\end{equation}
 which describes the scattering of an electron from an initial state
$\vert i\rangle=\vert0L\rangle$ in the left lead, to a final state
$\vert f\rangle=\vert N+1R\rangle$ in the right lead. Assuming that
the chemical potential difference between the to leads (electron reservoirs)
is $\mu_{R}=\mu_{L}+e{\cal V}$, with $e$ the modulus of the electron
charge and ${\cal V}$ the electromotive potential between the two
reservoirs, the transmission rate is given by \begin{eqnarray}
\frac{1}{\tau} & = & \frac{2\pi}{\hbar}\sum_{\alpha,\beta}f(E_{\alpha}^{L}-\mu_{L})[1-f(E_{\beta}^{R}-\mu_{R})]\times\nonumber \\
 &  & \vert T_{0L,N+1R}\vert^{2}\delta(E_{\alpha}^{L}-E_{\beta}^{R})\,,\end{eqnarray}
 where \begin{equation}
T_{0L,N+1R}=\langle0L\vert\hat{T}\vert N+1R\rangle=\langle0L\vert V\hat{G}V\vert N+1R\rangle\,,\end{equation}
 leading to a linear current \begin{eqnarray}
j=-\frac{e}{\tau} & = & {\cal V}\frac{2\pi e^{2}}{\hbar}\sum_{\alpha,\beta}(-)\left.\frac{\partial f(x)}{\partial x}\right\vert _{x=E_{\beta}-\mu_{L}}\times\nonumber \\
 &  & \vert T_{0L,N+1R}\vert^{2}\delta(E_{\alpha}^{L}-E_{\beta}^{R})\,.\end{eqnarray}
 As usual, the conductance is given by $g=j/{\cal V}$. At low temperature,
$g$ reads \begin{eqnarray}
g(\mu_{L}) & = & \frac{2\pi e^{2}}{\hbar}\vert T_{0L,N+1R}\vert^{2}\rho_{L}(\mu_{L})\rho_{R}(\mu_{L})\,\\
 & = & \frac{2\pi e^{2}}{\hbar}t_{R}^{2}t_{L}^{2}\vert G_{lr}^{1N}(\mu_{L})\vert^{2}\rho_{L}(\mu_{L})\rho_{R}(\mu_{L})\,.\label{g}\end{eqnarray}
 Naturally, the density of states $\rho_{L}(\mu_{L})$ and $\rho_{R}(\mu_{L})$
should be interpreted as the local density of states at sites 0 and
$N+1$, respectively. Equation (\ref{g}) is formally equal to that
derived by Caroli \textit{et al.}, \cite{caroli} using the Keldysh
formalism, and should be multiplied by a factor of two due to the
spin of the electrons.

We should stress here that our approach to the tunneling problem is
similar to that developed by Mujica \textit{et al.}, \cite{mujicaI,mujica,mujicabook}
but not exactly identical. These authors used the Newns coupling\cite{newns}
of the system to the leads, and their approach to the solution of
the Green's function is based on the properties of tridiagonal determinants
after using L\"owdin's partitioning technique,\cite{lowdin} whereas
we explicitly solve the Dyson's equation.


\subsection{Application to a three-sites quantum dot}

\label{gapllication}

\begin{figure}[ht]
 \includegraphics[width=7cm]{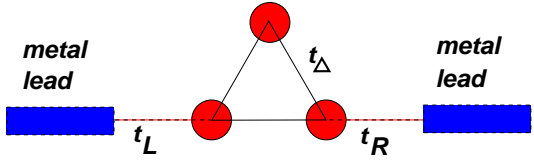} 

\caption{\label{Fig-triang} A three-sites quantum dot characterized by $t_{\triangle}$.
This is coupled to two metal leads by the couplings $t_{R}$ and $t_{L}$.}

\end{figure}

As an application of the formalism that can be worked out analytically
in full detail, we study the transport across a three-sites quantum
dot, as that shown in Fig. \ref{Fig-triang}. Although this is a very
simple model, it is relevant enough for the our illustrative purposes.

The Green's functions of the quantum dot depicted in Fig. \ref{Fig-triang}
has been computed in Eq. (\ref{Gtriangle}), for the case where the
coupling to the leads was neglected. Next, we need to compute Eq.
(\ref{G1N}) for this problem. To this end we need the surface Green's
functions $G_{00,L}$ and $G_{22,R}$, which lead to \begin{equation}
G_{00,L}^{ret}(E)=G_{22,R}^{ret}(E)=\frac{x}{\beta_{i}}-\frac{i}{\beta_{i}}\sqrt{1-x^{2}}\,,\label{Gsurf}\end{equation}
 with $x=E/(2\beta_{i})$, with $i=L,R$. Interesting enough, the
local density of states computed from (\ref{Gsurf}), does not diverge
at the band edge, as it happens with the density of states of an infinite
one-dimensional tight-binding model. \cite{economou} The calculation
of the matrix element of the $T-$matrix leads to

\begin{equation}
\vert G_{lr}^{1N}(\mu_{L})\vert^{2}=16t_{\triangle}^{2}\beta^{8}(t_{\triangle}+E-\epsilon_{0})^{2}{\cal D}^{-1}\,,\end{equation}
 with ${\cal D}$ given by \begin{eqnarray}
 &  & {\cal D}=4(4\beta^{2}-E^{2})\left[(t_{R}^{2}+t_{L}^{2})\beta^{2}(t_{\triangle}^{2}-(E-\epsilon_{0})^{2})\right.\nonumber \\
 &  & \left.+t_{R}^{2}t_{L}^{2}E(E-\epsilon_{0})\right]^{2}+\left[2E\beta^{2}(t_{L}^{2}+t_{R}^{2})(-t_{\triangle}^{2}+(E-\epsilon_{0})^{2})\right.\nonumber \\
 &  & -2t_{L}^{2}t_{R}^{2}(E^{2}-2\beta^{2})(E-\epsilon_{0})\nonumber \\
 &  & \left.+4\beta^{4}(t_{\triangle}+E-\epsilon_{0})^{2}(2t_{\triangle}-E+\epsilon_{0})\right]^{2}\,.\label{Dcal}\end{eqnarray}
 The algebraic form of ${\cal D}$ suggests that for an array of quantum
dots, giving a simple analytical form as Eq. (\ref{G1N}) may not
be possible in general, and the last steps of a given particular calculation
may have to be done numerically.


\subsection{Defects}

\label{deff}

In this section we describe the effect of defects on the electronic
spectrum of the array of dots as well as on its transport properties.
Let us again consider the generic situation described in Fig. \ref{Fig_array},
and consider as a simple and specific example that at the site $(x,l)$
($1\le x\le N$) there is an adatom. We want to study what is the
effect of this adatom on the spectrum of the system and latter on
its transport properties. A particular study of the effect of an adatom
on the conductance of a quantum wire was done by Kwapi\'{n}ski. \cite{kwapinski}
The presence of the adatom adds an extra term to the Hamiltonian (\ref{HD})
of the form \begin{equation}
H_{AD}=H_{AD}^{0}+V_{AD}\,,\end{equation}
 with \begin{subequations} \begin{eqnarray}
H_{AD}^{0} & = & \epsilon_{\odot}\vert AD\rangle\langle AD\vert\,,\\
V_{AD} & = & t_{\odot}(\vert AD\rangle\langle xl\vert+\vert xl\rangle\langle AD\vert)\end{eqnarray}
 \end{subequations} where $\vert AD\rangle$ is the electronic state
in the adatom, $t_{\odot}$ is the electronic hopping between the
adatom and the $(x,l)$ site of the array, and $\epsilon_{\odot}$
is the local electronic energy in the adatom.

It is again straightforward to apply the Dyson equation formalism
to compute the exact Green's function in the presence of the impurity,
leading to

\begin{equation}
\langle n\alpha\vert\hat{G}\vert m\beta\rangle=\langle n\alpha\vert\hat{G}^{0}\vert m\beta\rangle+\langle n\alpha\vert\hat{G}^{0}\vert xl\rangle T\langle xl\vert\hat{G}^{0}\vert m\beta\rangle\,,\label{Gimpu}\end{equation}
 with the $T$ matrix given by \begin{equation}
T=t_{\odot}^{2}[z-\epsilon_{\odot}-t_{\odot}^{2}\langle xl\vert\hat{G}^{0}\vert xl\rangle]^{-1}\,,\end{equation}
 and the matrix elements $\langle n\alpha\vert\hat{G}^{0}\vert m\beta\rangle$
are computed from the resolvent $\hat{G}^{0}=(z-H_{T})^{-1}$, with
$H_{T}$ defined by Eq. (\ref{HD}). From Equation (\ref{Gimpu})
we can compute Eq. (\ref{G1N}) and from this the conductance given
by Eq. (\ref{g}).

In addition, we can compute the Green function of the impurity, determining
how the energy level is modified due to the coupling to the bath of
electrons propagating along the array. This is given by ($G_{AD}\equiv\langle AD\vert\hat{G}\vert AD\rangle$)
\begin{equation}
G_{AD}=[z-\epsilon_{\odot}-t_{\odot}^{2}\langle xl\vert\hat{G}^{0}\vert xl\rangle]^{-1}\,.\label{GAD}\end{equation}
 From Equation (\ref{GAD}), the local density of states at the impurity
can be computed as $\rho_{AD}=-1/\pi\Im G_{AD}^{ret}$. The accepted
$\theta$ values are now the solution of $1/G_{AD}=0$.


\subsection{Defects: an application}

\begin{figure}[ht]
 \includegraphics[width=7cm]{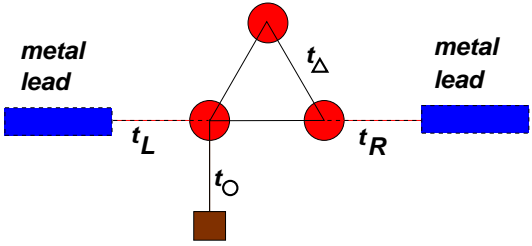} 

\caption{\label{Fig_deff} A triangular dot with the atoms coupled together
by $t_{\triangle}$, which in turn is coupled to two metal leads
by the coupling constants $t_{R}$ and $t_{L}$. An add atom (represented by
the square) is connected to one of the atoms of the dot via the hopping
$t_{\odot}$.}

\end{figure}

Again we make a simple application of the formalism of Sec.
\ref{deff} considering the system depicted in Fig. \ref{Fig_deff}.

The $G^{0}$ Green's functions entering in Eq. (\ref{G1N}) are given
in this example by \begin{subequations} \label{G0adatom} \begin{eqnarray}
G_{ll}^{0} & = & (\lambda-1/2)(1+2\lambda)(2t_{\triangle}\lambda+\Delta\epsilon){\cal D}^{-1}\,,\\
G_{rr}^{0} & = & [(\lambda-1/2)(1+2\lambda)(2t_{\triangle}\lambda+\Delta\epsilon)\nonumber \\
 & - & \lambda t_{\odot}^{2}/t_{\triangle}]{\cal D}^{-1}\,,\\
G_{lr}^{0} & = & -(2t_{\triangle}\lambda+\Delta\epsilon){\cal D}^{-1}\,,\end{eqnarray}
 \end{subequations} with ${\cal D}$ given by \begin{equation}
{\cal D}=(\lambda-1/2)[2t_{\triangle}(-1+\lambda+2\lambda^{2})(2t_{\triangle}\lambda+\Delta\epsilon)-t_{\odot}^{2}(1+2\lambda)]\,,\end{equation}
 with $\Delta\epsilon=\epsilon_{0}-\epsilon_{\odot}$. It is interesting
to note that the eigenvalue $\lambda=1/2$ of the non-perturbed triangular
dot is not modified by the adatom. The calculation of the conductance
is now just a matter of using Eq. (\ref{G0adatom}) in Eq. (\ref{G1N}).


\section{Discussion}

We have presented in full detail a method to compute the lattice
Green's functions of an array of quantum dots for the cases when the
array is isolated as well as when it is coupled to two metallic
leads. The effect of the leads is to produce a self-energy which has
both a real ($\Re\Sigma$) and an imaginary ($\Im\Sigma$) parts. In the
case of a single quantum dot, $\Re\Sigma$ will renormalize the energy
levels in the dot whereas $\Im\Sigma$ makes the energy levels
non-stationary. Both terms contribute to the charge transport through
the dot.

The formalism is general and flexible enough to allow for the study
of how localized defects affect both the energy spectrum and the transport
properties. We can consider both the case when the defect acts as
local potential, i.e., diagonal disorder, and when the defect changes locally
the values of the hopping integrals, i.e., off-diagonal disorder. Also more
than one defect can be attached to the quantum dot array, at different
positions in the lattice. In the case of a random distribution of
impurities an approximate treatment such as the CPA\cite{economou}
can be used to compute the full lattice Green's function self-consistently.\cite{peres0}

The generalization of the present approach to two dimensions should
present no difficulties, allowing for the possibility to proceed analytically
in the calculation of the energy spectrum and transport properties
of finite size two-dimensional ribbons. This possibility will be explored
in a forthcoming publication. The more restrictive aspect of the method
could be related to the calculation of the allowed values of $\theta$,
since these are to be computed at the same time the values of the
energy eigenvalues are determined. For an array of $N$ dots, each
dot having $N_{s}$ sites, the determination of the spectrum following
a brute force approach would require the diagonalization of a matrix
of dimension $(N_{s}\times N)^{2}$. In our approach this is reduced
to the determination of the zeros of a polynomial of degree $N_{s}$.
If we consider the case of periodic boundary conditions, the $\theta$
values are given by $\theta_{\ell}=2\pi\ell/N$, with $\ell=0,\pm1,\pm2,\ldots N/2$
(assuming $N$ even), and the equation giving the energy spectrum
is the same which we would have obtained if we had done a Fourier transform
in the initial Hamiltonian.

It important to stress that our approach can easily include the
case where the quantum dot is represented by a continuous model. In
this case the Green's function of the dot has the form $G(\mathbf{r},\mathbf{r}',E)$,
where $\mathbf{r}$ and $\mathbf{r}'$ are two-dimensional vectors
characterizing the position in the dot. In order to apply the developed
formalism we only need to choose the values of $\mathbf{r}$ to which
the dots connect among themselves.

Finally we note that our description of the transport is easily generalized to include
finite values of the potential bias between the leads. In this case, an appropriate
treatment of the problem requires to solve for the Green's function together with
an iterative solution of the Poisson's equation. For quasi-one-dimensional systems this
does not require powerful computational facilities. This will be addressed in a forthcoming
publication.


\section*{Acknowledgements}

This work was supported by the ESF Science Programme INSTANS 2005-2010,
and by FCT under the grant PTDC/FIS/64404/2006. NMRP thanks JL Ribeiro and
JMP Carmelo for suggestions and support.


\appendix

\section{An alternative solution of the finite chain problem I}

\label{apJLS}

We develop here an approach to the solution of the finite chain problem
that builds the symmetries between the two coordinates of the Green
function form the start. This method can also be used to tackle the
more general problem of the quantum dot array. In terms of the resolvents
and of $V$, Dyson's equation can take two alternative forms,

\begin{subequations} \label{dyson:app}\begin{eqnarray}
\hat{G} & = & \hat{G}^{0}+\hat{G}^{0}V\hat{G}\,,\label{dysona}\\
\hat{G} & = & \hat{G}^{0}+\hat{G}V\hat{G}^{0}\,.\label{eq:dysonb}\end{eqnarray}
 \end{subequations} Forming matrix elements with the basis state
vectors and using $\hat{\mathbf{1}}=\sum_{n=1}^{N}\vert n\rangle\langle n\vert$,
one obtains \begin{subequations} \label{eq:matrixG}\label{matrix} \begin{eqnarray}
\langle n\vert\hat{G}\vert m\rangle & = & G^{0}\delta_{nm}\nonumber \\
 & - & tG^{0}(\langle n-1\vert\hat{G}\vert m\rangle+\langle n+1\vert\hat{G}\vert m\rangle)\,,\label{eq:matrixG1}\\
\langle n\vert\hat{G}\vert m\rangle & = & G^{0}\delta_{nm}\nonumber \\
 & - & tG^{0}(\langle n\vert\hat{G}\vert m-1\rangle+\langle n\vert\hat{G}\vert m+1\rangle)\,,\label{eq:matrixG2}\end{eqnarray}
 \end{subequations} with $G^{0}=(z-\epsilon_{0})^{-1}$. By taking
the sum and the difference of these equations, one derives equivalent
conditions which are more symmetrical in the two coordinates of the
Green function,

\begin{subequations} \label{matrixsum_dif} \begin{eqnarray}
\frac{1}{G^{0}}\langle n\vert\hat{G}\vert m\rangle & = & \delta_{nm}-\frac{t}{2}\left(\langle n-1\vert\hat{G}\vert m\rangle+\langle n+1\vert\hat{G}\vert m\rangle\right.\nonumber \\
 & + & \left.\langle n\vert\hat{G}\vert m-1\rangle+\langle n\vert\hat{G}\vert m+1\rangle\right)\,,\label{eq:matrixsumG}\\
0 & = & \langle n-1\vert\hat{G}\vert m\rangle+\langle n+1\vert\hat{G}\vert m\rangle\nonumber \\
 & - & \langle n\vert\hat{G}\vert m+1\rangle-\langle n\vert\hat{G}\vert m+1\rangle.\label{eq:matrixdifG}\end{eqnarray}
 \end{subequations} Apart from the Kronecker delta term $\delta_{nm}$, Eq. (\ref{eq:matrixsumG})
defines the wavefunction of two particles with a tight-binding Hamiltonian
and an eigenvalue $1/G^{0}$. It is obvious that the difference of
two solutions of this set of equations will be a solution of the corresponding
homogeneous system (without the $\delta_{nm}$ term). Our strategy
for finding $\langle n\vert\hat{G}\vert m\rangle\equiv G_{nm}$ is
the following: (i) we construct the general solution of the homogeneous
system assuming plane waves in the two {}``particle'' coordinates;
(ii) we find \emph{one} solution of the full non-homogeneous system,
by allowing different plane wave solution for $n\leq m$ and $n>m$,
very much in the spirit of the Bethe solution for two particles with
local interactions, moving in one dimension; (iii) the general solution
is just the particular solution of (ii) added to a general linear
combinations of the solutions of (i). To determine the latter we require
additional boundary conditions, which, for a finite chain are:

\begin{subequations} \label{bcG} \begin{eqnarray}
G_{0m} & = & 0\,,\label{eq:bc1G}\\
G_{N+1m} & = & 0\,,\label{eq:bc2G}\\
G_{n0} & = & 0\,,\label{eq:bc3G}\\
G_{nN+1} & = & 0\,.\label{eq:bc4G}\end{eqnarray}
 \end{subequations} We begin by writing a general solution of the
homogeneous system in the form \begin{equation}
\psi_{nm}=Ae^{i\theta_{1}n+i\theta_{2}m}.\end{equation}
 Inserting this trial solution in Eq. (\ref{eq:matrixdifG}) we get
the condition $\cos\theta_{1}-\cos\theta_{2}=0$; we must have $\theta_{1}=\pm\theta_{2}\equiv\theta$.
With this condition, $\psi_{nm}$ is a solution of the homogeneous
version of Eq. (\ref{eq:matrixsumG}) provided

\begin{equation}
\frac{1}{G^{0}}=-2t\cos\theta\,.\label{GtetaII}\end{equation}
 So the solution of the homogeneous equations is a linear superposition
of waves \begin{equation}
\psi_{nm}(\theta)=Ae^{i\theta(n-m)}+Be^{i\theta(n+m)},\label{eq:free_sol}\end{equation}
 where $\theta$ solves Eq. (\ref{GtetaII}).

We now address the determination of one solution of the full non-homogeneous
Eqs. (\ref{matrixsum_dif}), which we write in the form $\phi_{nm}=\psi_{nm}^{<}$,
for $n\leq m$, and $\phi_{nm}=\psi_{nm}^{>}$, for $n>m$, where
$\psi_{nm}^{<}$ and $\psi_{nm}^{>}$ are two different solutions
of the homogeneous system. There are only two conditions that mix
$\psi^{<}$ and $\psi^{>}$, namely

\begin{eqnarray*}
\frac{1}{G^{0}}\psi_{nn}^{<} & = & 1-\frac{t}{2}\left(\psi_{n+1n}^{>}+\psi_{n-1n}^{<}+\psi_{nn+1}^{<}+\psi_{nn-1}^{>}\right)\,,\\
\frac{1}{G^{0}}\psi_{n+1n}^{>} & = & -\frac{t}{2}\left(\psi_{n+2n}^{>}+\psi_{nn}^{<}+\psi_{n+1n+1}^{<}+\psi_{n+1n-1}^{>}\right).\end{eqnarray*}
 Because $\psi^{<}$ and $\psi^{>}$ are solutions of the homogeneous
system, we can easily transform these conditions into

\begin{subequations} \label{bc_eqs} \begin{eqnarray}
1 & = & \frac{t}{2}\left(\psi_{n+1n}^{>}-\psi_{n+1n}^{<}+\psi_{nn-1}^{>}-\psi_{nn-1}^{<}\right)\,,\label{eq:bceq_1}\\
0 & = & \psi_{nn}^{<}-\psi_{nn}^{>}+\psi_{n+1n+1}^{<}-\psi_{n+1n+1}^{>}\,.\label{eq:bc_eq_2}\end{eqnarray}
 \end{subequations} These conditions cannot be fulfilled by solutions
which are function of $n+m$, so we must have: \begin{eqnarray*}
\psi_{nm}^{<} & = & A^{<}e^{i\theta(n-m)}+B^{<}e^{-i\theta(n-m)}\,,\\
\psi_{nm}^{>} & = & A^{>}e^{i\theta(n-m)}+B^{>}e^{-i\theta(n-m)}\,.\end{eqnarray*}
 Inserting these trial functions in Eqs. (\ref{bc_eqs}) and solving
the corresponding linear equations for the constants, one gets, \[
A^{<}=-B^{<}=-A^{>}=B^{>}=-\frac{1}{4it\sin\theta},\]
 leading to a solution \[
\phi_{nm}=\frac{1}{2t\sin\theta}\sin\left(\theta\left|n-m\right|\right),\]
exactly as found in section \ref{sub:The-single-chain}, eq,(\ref{eq:particular_sol})
We have now carried out points (i) and (ii) outlined above, and obtained
the general solution of Eqs. (\ref{matrixsum_dif}), as\[
G_{nm}=\psi_{nm}+\frac{\sin\left(\theta\left|n-m\right|\right)}{2t\sin\theta}\,,\]
 where $\psi_{nm}$ is superposition of waves of the form (\ref{eq:free_sol})
with $\theta$ satisfying Eq. (\ref{GtetaII}). To enforce the boundary
conditions it proves more convenient to write the solution in sines
and cosines as \begin{eqnarray*}
G_{nm} & = & A\cos\left[\theta\left(n-m\right)\right]+B\sin\left[\theta\left(n-m\right)\right]\\
 & + & C\cos\left[\theta\left(n+m\right)\right]+D\sin\left[\theta\left(n+m\right)\right]\\
 & + & \frac{\sin\left(\theta\left|n-m\right|\right)}{2t\sin\theta}\,.\end{eqnarray*}
 To derive the values of these constants, we insert this solution
in the Eqs. (\ref{bcG}), use the linear independence of the sine
and cosine functions and arrive at the final result:

\begin{eqnarray}
G_{nm}(z) & = & \frac{1}{2t}\frac{\cos\left[\theta\left(N+1\right)\right]}{\sin\left[\theta\left(N+1\right)\right]\sin\theta}\nonumber \\
 & \times & \left\{ \cos\left[\theta\left(n-m\right)\right]-\cos\left[\theta\left(n+m\right)\right]\right\} \nonumber \\
 & - & \frac{1}{2t}\left\{ \frac{\sin\left[\theta\left(n+m\right)\right]}{\sin\theta}-\frac{\sin\left[\theta\left|n-m\right|\right]}{\sin\theta}\right\} \,,\label{eq:Gsol2}\end{eqnarray}
 which is the same solution as Eq. (\ref{Gsol}).

\section{An alternative solution of the finite chain problem II}

\label{apTS}
In the previous appendix, the Dyson equation was written in two
alternative forms, see Eqs. (\ref{dyson:app}) and (\ref{eq:matrixG}).
 
In Eq. (\ref{eq:matrixG1}), the ket is unchanged and it can be thought of as a tight-binding equation for the bra of $\langle n|G|m\rangle$ with an inhomogeneity at site $m$. Since we are dealing with a real Hamiltonian, we can make the following ansatz for $G_{n,m}=\langle n|G|m\rangle$:
\begin{align}
G_{nm}=
\left\{ \begin{array}{cl} G_{nm}^<=a_1(m)\cos \theta n+a_2(m)\sin \theta n&\;,\;n<m\\
    G_{nm}^>=b_1(m)\cos \theta n+b_2(m)\sin \theta n&\;,\;n\geq m
\end{array} \right.
\label{ansatz}
\end{align}
where $a_i(m)$ and $b_i(m)$ ($i=1,2$) are arbitrary functions of $m$.

In Eq. (\ref{eq:matrixG2}), the bra is unchanged and it can be thought of as a tight-binding equation for the ket of $\langle n|G|m\rangle$ with an inhomogeneity at site $n$. We can thus make the following ansatz:
\begin{align}
G_{nm}=
\left\{ \begin{array}{cl} G_{nm}^<=c_1(n)\cos \theta m+c_2(n)\sin \theta m&\;,\;m>n\\
    G_{nm}^>=d_1(n)\cos \theta m+d_2(n)\sin \theta m&\;,\;m\leq n
\end{array} \right.
\label{ansatzB}
\end{align}
where $c_i(n)$ and $d_i(n)$ ($i=1,2$) are arbitrary functions of $n$.

Combining Eqs. (\ref{ansatz}) and (\ref{ansatzB}), we arrive to the following ansatz for the Green's function:
\begin{widetext}
\begin{align}
G_{nm}=
\left\{ \begin{array}{cl} G_{n,m}^<=a_1\cos\theta n\cos \theta m+a_2\cos\theta n\sin \theta m+a_3\sin\theta n\cos \theta m+a_4\sin\theta n\sin \theta m&\;,\;n<m\\
    G_{nm}^>=b_1\cos\theta n\cos \theta m+b_2\cos\theta n\sin \theta m+b_3\sin\theta n\cos \theta m+b_4\sin\theta n\sin \theta m&\;,\;n\geq m
\end{array} \right.
\label{ansatzCombined}
\end{align}
\end{widetext}
Notice that now the coefficients are site-independent.

From the boundary conditions $G_{0m}^<=G_{n0}^>=0$, we obtain $a_1=a_2=b_1=b_3=0$. From the boundary conditions $G_{N+1m}^<=G_{nN+1}^>=0$, we have $a_3=-a_4\tan\theta(N+1)$ and $b_2=-b_4\tan\theta(N+1)$. The matching condition at $n=m-1$ yields $G_{mm}^<=G_{mm}^>$ (continuity of the Green's function) and thus $a_4=b_4$. The matching condition at $n=m$ yields $G_{m-1m}^<-G_{m-1m}^>=t^{-1}$ (discontinuity of the derivative of the Green's function) and thus $a_4=(\sin\theta\tan\theta(N+1)t)^{-1}$. The final result can therefore be written as
\begin{align}
G_{nm}^<&=\frac{1}{t}\frac{-\tan\theta(N+1)\sin\theta n\cos\theta m+\sin\theta n\sin\theta m}{\tan\theta(N+1)\sin\theta}\notag\\
&=-\frac{1}{t}\frac{\sin\theta(N+1-m)\sin\theta n}{\sin\theta(N+1)\sin\theta}\;,\\
G_{nm}^>&=G_{mn}^<\;.
\end{align}
Again $\theta$ is determined by the dispersion relation Eq. (\ref{GtetaII}). This yields an alternative (but equivalent) representation of the Green's functions of a tight-binding chain with open boundaries.

The extension to the more general case is analogous, but one has to take special care by defining the matching conditions because the unperturbed Green's function is now a matrix. It then follows that the Green's function for the non-diagonal matrix elements which are not constrained by the boundary conditions will be discontinuous for energies which are not eigenenergies of the unperturbed system. 


\section{Matrix $M$ of Eq. (\ref{Mvb}) and matrix $V$ of Eq. (\ref{N})}

\label{apI} 
The matrix $M$ of Eq. (\ref{Mvb}) is given by \begin{equation}
M=\left[\begin{array}{ccc}
L_{0}+tG_{lr}^{0}L_{1} & tG_{ll}^{0}L_{2} & 0\\
0 & tG_{rr}^{0}L_{4} & L_{5}+tG_{lr}^{0}L_{3}\\
tG_{rr}^{0}L_{4} & tG_{lr}^{0}(L_{2}-L_{4}) & -tG_{ll}^{0}L_{3}\end{array}\right]\,,\end{equation}
 with the functions $L_{i}$, with $i=0,\ldots,5$ given by \begin{subequations}
\begin{eqnarray}
L_{0} & = & \cos(m\theta)-\cot(N\theta+\theta)\sin(m\theta)\,,\\
L_{1} & = & \cos(m\theta+\theta)-\cot(N\theta+\theta)\sin(m\theta+\theta)\,,\\
L_{2} & = & [\cot(m\theta)-\cot(N\theta+\theta)]\sin(m\theta-\theta)\,,\\
L_{3} & = & \sin(m\theta-\theta)\,,\\
L_{4} & = & \cos(m\theta+\theta)-\cot(N\theta+\theta)\sin(m\theta+\theta)\,,\\
L_{5} & = & \sin(m\theta)\,.\end{eqnarray}
 \end{subequations} 

The matrix $V$ of Eq. (\ref{N}) is given by
\begin{equation}
V=\left[\begin{array}{ccc}
\cos(m\theta) & \sin(m\theta) & 0\\
\cos(m\theta-\theta)& \sin(m\theta-\theta) & 0\\
\cos(m\theta+\theta) & a(\theta) & b(\theta)
\end{array}\right]\,,
\end{equation}
 with the functions $a(\theta)$ and $b(\theta)$ given by 
\begin{subequations}
\begin{eqnarray}
a(\theta)&=&\cos(m\theta+\theta)\tan(m\theta)\,,\\
b(\theta)&=& \sin(m\theta+\theta)-\cos(m\theta+\theta)\tan(m\theta)\,.
\end{eqnarray}
\end{subequations}
The vector $\mathbf{q}^T=[q_1,q_2,q_3]$ entering in Eq. (\ref{N})
has its components given by
\begin{subequations}
\begin{eqnarray}
q_1&=&(tP_L)^{-1}[tG^0_{rl}P_L-t^3G^0_{rl}G^0_{rr}G^0_{ll}\sin(m\theta-\theta)\nonumber\\
&\times&\sin(N\theta+\theta-m\theta)\nonumber\\
&+& t^2G^0_{rr}G^0_{ll}\sin(N\theta-m\theta)\tilde P_{m-1})]\,,\\
q_2&=&(tP_L)^{-1}[-P_L+t^2G^0_{rr}G^0_{ll}\sin(N\theta+\theta-m\theta)\nonumber\\
&\times&\sin(m\theta-\theta)\,,\\
q_3&=&(tP_L)^{-1}[-P_L+t^2G^0_{rr}G^0_{ll}\sin(N\theta-m\theta)\nonumber\\
&\times&\sin(m\theta)\,.
\end{eqnarray}
\end{subequations}


\section{Full analytical expressions for the Green's functions}

\label{apII}
After using the boundary conditions and three of the four time reversal conditions, the
Ansatz for the Green's functions for $n<m$ is
\begin{subequations}
\begin{eqnarray}
 G^{nm}_{ll}&=&A^<_{ll}[\cos(n\theta)-\cot(m\theta)\sin(n\theta)]\nonumber\\
&+&A^>_{ll}[\cot(m\theta)-\cot(N\theta+\theta)]\sin(n\theta)\,,\\
G^{nm}_{rl}&=&A^>_{lr}[\cot(m\theta)-\cot(N\theta+\theta)]\sin(n\theta)\,,\\
G^{nm}_{lr}&=&A^<_{lr}\cos(n\theta)+B^<_{lr}\sin(n\theta)\,,\\
G^{nm}_{rr}&=&B^<_{rr}\sin(n\theta)\,,
\end{eqnarray}
\end{subequations}
and for $n>m$ is 
\begin{subequations}
\begin{eqnarray}
 G^{nm}_{ll}&=&A^>_{ll}[\cos(n\theta)-\cot(N\theta+\theta)\sin(n\theta)]\,,\nonumber\\
G^{nm}_{rl}&=&A^>_{rl}\cos(n\theta)+B^>_{rl}\sin(n\theta)\,,\\
G^{nm}_{lr}&=&A^>_{lr}[\cos(n\theta)-\cot(L\theta+\theta)\sin(n\theta)]\,,\\
G^{nm}_{rr}&=&A^>_{rr}[\cos(n\theta)-\cot(m\theta)\sin(n\theta)]\nonumber\\
&+&B^<_{rr}\sin(n\theta)\,.
\end{eqnarray}
\end{subequations}

Following the method described in the bulk of the paper, 
the full analytical expressions for the Green's functions are given
by

\begin{eqnarray}
G_{rr}^{ij}=\left\{ \begin{array}{c}
-\frac{G_{rr}^{0}}{P_{L}}\sin(i\theta)\tilde{P}_{L-j},\hspace{0.5cm}i<j\,,\\
-\frac{G_{rr}^{0}}{P_{L}}\sin(j\theta)\tilde{P}_{L-i},\hspace{0.5cm}i>j\,,\end{array}\right.\end{eqnarray}

\begin{eqnarray}
G_{ll}^{ij}=\left\{ \begin{array}{c}
-\frac{G_{ll}^{0}}{P_{L}}\sin[(N+1-j)\theta]\tilde{P}_{i-1},\hspace{0.5cm}i<j\,,\\
-\frac{G_{ll}^{0}}{P_{L}}\sin[(N+1-i)\theta]\tilde{P}_{j-1},\hspace{0.5cm}i>j\,,\end{array}\right.\end{eqnarray}

\begin{eqnarray}
G_{rl}^{ij}=\left\{ \begin{array}{c}
\frac{tG_{rr}^{0}G_{ll}^{0}}{P_{L}}\sin(i\theta)\sin[(N+1-j)\theta],\hspace{0.5cm}i<j\,,\\
\frac{1}{tP_{L}}\tilde P_{L-i}\tilde P_{j-1}\hspace{0.5cm}i\ge j\,,\end{array}\right.\end{eqnarray}
 and \begin{eqnarray}
G_{lr}^{ij}=\left\{ \begin{array}{c}
\frac{1}{tP_{L}}\tilde P_{L-j}\tilde P_{i-1}\hspace{0.5cm}i\le j\,,\\
\frac{tG_{rr}^{0}G_{ll}^{0}}{P_{L}}\sin(j\theta)\sin[(N+1-i)\theta],\hspace{0.5cm}i>j\,.\end{array}\right.\end{eqnarray}
Note that $G_{lr}^{ij}=G_{rl}^{ji}$ and that the diagonal Green's functions obey
\begin{eqnarray}
G_{ll}^{>,i+1i}-G_{ll}^{<,i+1i}=\frac{G_{ll}^0}{tG_{lr}^0}\,,\\
G_{rr}^{<,i-1i}-G_{rr}^{>,i-1i}=\frac{G_{rr}^0}{tG_{rl}^0}\,,
\end{eqnarray}
which is similar to Eqs. (\ref{eq:bc2}) and (\ref{GderivI}), which is the generalization to the lattice
of the discontinuity of the first derivative of a Green's function.

Similar results to those given in this Appendix have been also obtained in Ref. [\onlinecite{onipkof}]
in the context of organic molecular systems,
but  no hints about the method
used to derive them was given.


\end{document}